\documentclass[graybox]{svmult}

\usepackage{type1cm}        
\usepackage{makeidx}         
\usepackage{graphicx}        
\usepackage{multicol}        
\usepackage[bottom]{footmisc}

\usepackage{newtxtext}       %
\usepackage{newtxmath}       

\makeindex             

\begin{document}

\title*{Space Debris - Optical Measurements}
\author{Ji\v{r}\'{i} \v{S}ilha}
\institute{Ji\v{r}\'{i} \v{S}ilha \at Comenius University, Faculty of Mathematics, Physics and Informatics, 84248 Bratislava, Slovakia, \email{silha1@uniba.sk}}
%
%
\maketitle

\abstract{Space debris is a major threat to the satellite infrastructure. A collision with even small particle, e.g. 1 cm of size, can cause a catastrophic event when the parent body, spacecraft or upper stage, will break up into hundreds of trackable fragments. Space debris research helps to discover, monitor and characterize these objects, identify their origin and support their active removal. Surveys with optical telescopes aim to discover new objects for cataloguing and to increase the accuracy of space debris population models. The follow-up observations are performed to improve their orbits or to investigate their physical characteristics. We will present the space debris population, its orbital and physical characteristics and we will discuss the role which the optical telescopes play in space debris research. We will also discuss the adopted astronomical techniques like astrometry, photometry and spectroscopy used in the space debris domain.\newline\indent}

\section{Space debris}
\label{sec:debris}

Space debris, also known as orbital debris, can be defined as man-made object which is situated on geocentric orbit and have no longer any purpose. There are many sources and types of space debris with different origins, trajectories and physical parameters. 

\subsection{Satellite infrastructure}
\label{subsec:sat_inf}
Since the first launch of Sputnik 1 in 1957 thousands of satellites have been put on orbit around the Earth. They fulfill various tasks, from telecommunication, trough scientific and meteorological missions, and from military support, to broadband internet coverage. Additionally, the continuous satellite launches, the long exposure of hardware to the harsh space environment, and fragmentations of satellites and upper stages led to an unwanted population of debris objects orbiting the earth.

\subsection{Spatial distribution}
\label{subsec:spatial}
\begin{figure}[b]
\sidecaption
\includegraphics[width=12.0cm,angle=0]{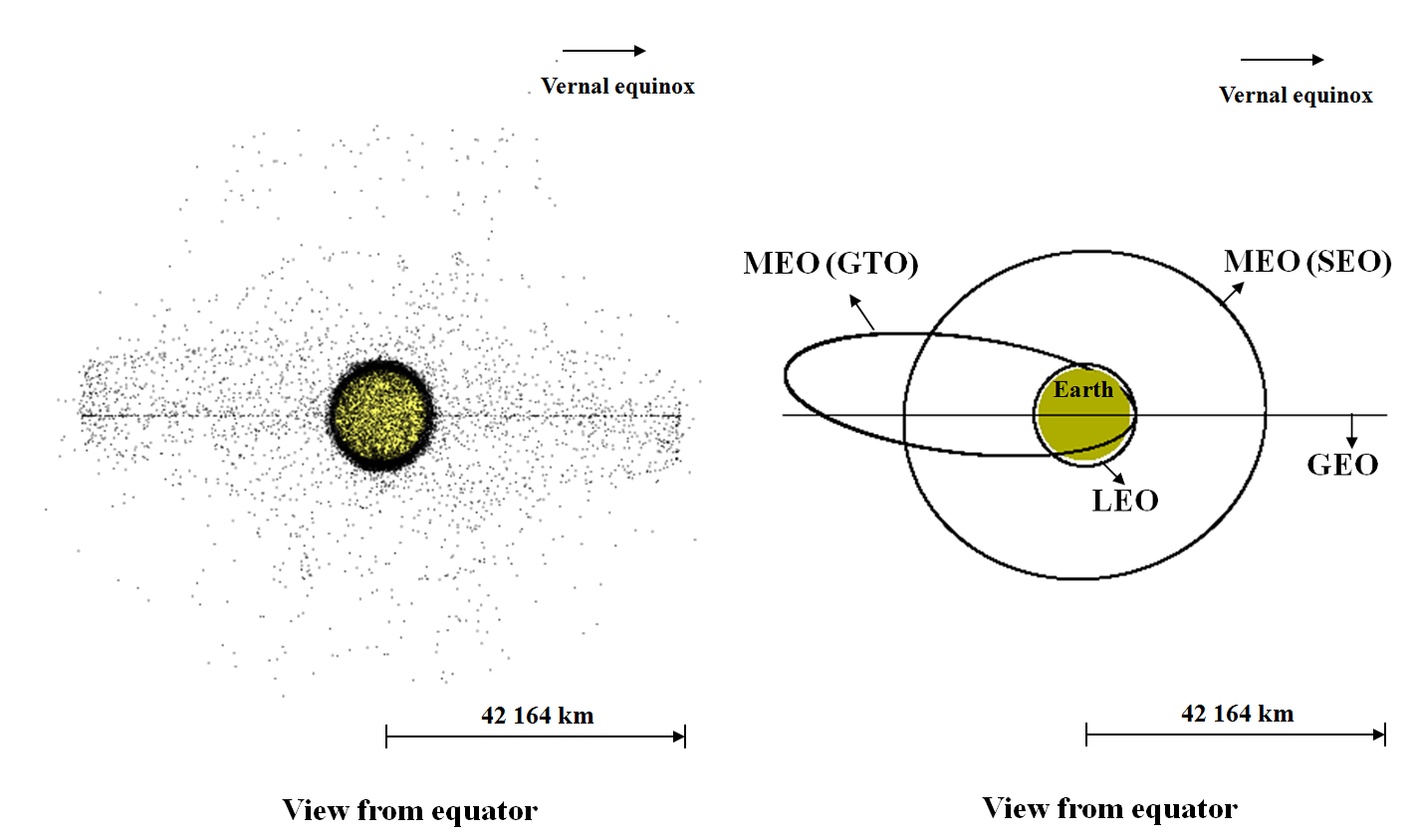}
%
%
\caption{Space debris situation as of January 2019. The line of sight is in the equator and perpendicular to the vernal equinox direction. Data were obtained from \cite{Spacetrack}.}
\label{fig:spatial}       
\end{figure}

The orbital distribution of catalogued debris is directly associated with the operational orbits of satellites. A snapshot of the space debris spatial distribution as of January 2019 can be seen in Figure \ref{fig:spatial}, rendered by using data from the US public catalogue \cite{Spacetrack}. As seen in Figure \ref{fig:spatial}, there are several types of geocentric orbits. They can be classified according to their orbital elements.

%

There are several slightly different definitions of orbital regimes in literature. The following definitions should be considered as more generic, but they follow the general consensus. The most populated regime is the low earth orbit (LEO) with mean altitude above the earth surface lower than 2,000 km, which corresponds to orbital periods of $P$ $\textless$ 2.2 hours. In general, LEOs have small orbit eccentricities ($e$ $\textless$ 0.1) and inclinations of $\textless$ 100$^{\circ}$. According to the public catalogue, almost 80 $\%$ of all catalogued objects are located on LEO, with most of them fragments. In Figure \ref{fig:spatial} the LEO population is represented by the dense region closely surrounding the earth. 

A very unique type of orbit is the geosynchronous earth orbit (GEO). Its orbits have periods close to 24 hours, inclinations between 0$^{\circ}$ and 15$^{\circ}$ and eccentricities close to 0. Specifically, truly geostationary orbits are near-circular with inclinations close to 0$^{\circ}$ and orbital periods of one sidereal day (23 hours, 56 minutes and 4.1 seconds), with mean altitudes above the earth surface of 35,786 km. Spacecraft located on this type of orbit remain above the same longitude in the equatorial plane. According to the public catalogue about 7 $\%$ of objects are in or on GEO. They are mostly payloads and rocket bodies. In Figure \ref{fig:spatial} the GEO population is represented by the slightly dispersed ring with orbit radii around 42,000 km and inclinations up to 15$^{\circ}$. 

Objects on medium earth orbits (MEO) have periods between 2.2 hours and 24 hours, and eccentricities covering wide ranges. Part of MEO are semi-synchronous orbits (SEO). These are used for navigation systems such as the US Global Position System (GPS), the Russian Globalnaya navigatsionnaya sputnikovaya sistema (GLONASS), and the European Galileo navigation system. These are sometimes referred to as Global Navigation Satellite Systems (GNSS). SEOs have periods around 12 hours and are near-circular ($e$ close to 0). Very common are eccentric MEO orbits known as geosynchronous transfer orbits (GTO) ($i$ mostly below 30$^{\circ}$) and Molniya orbits ($i$ between 60$^{\circ}$ and 70$^{\circ}$). About 8 $\%$ of all catalogued objects are located in the MEO regime. It is populated by satellites, rocket bodies, fragmentation debris and mission-related objects. In Figure \ref{fig:spatial} the MEO population can be seen between the LEO and GEO regions.  


Objects orbiting earth above GEO altitudes make only a small fraction of all catalogued objects. They are usually rocket bodies and science missions. Their orbits can be denoted as high elliptic orbits (HEO), or super GEO. 

%
%
%

\subsection{Origins and sources}
\label{subsec:origins}

The catalog orbit population is mainly composed of space debris. The largest and also easiest to track are non-functional payloads and spent upper stages of rocket bodies (R/B). More than 97 $\%$ of the total mass located on earth orbits is concentrated in this type of debris, along with functional spacecraft \cite{LIOU11} (the mass of the International Space Station (ISS) is not included here). The most abundant objects are roughly larger than 10 cm. They are fragments from payloads and rocket bodies, shortly denotaed as fragmentation debris. Fragmentations can be caused by different mechanisms \cite{BRAUN2017}, e.g., explosions, collisions \cite{PARDINI2011557}, intentional destructions \cite{LIOU20091407}, or malfunctions. Shapes, sizes and material types of fragments differ for every piece. During the spacecraft launching process many other additional objects can be released. Protective covers, launch adapters and objects lost by astronauts are part of debris called mission-related objects (MRO) \cite{JOHNSON2008}.

\begin{figure}[b]
\sidecaption
\includegraphics[width=7.5cm,angle=0]{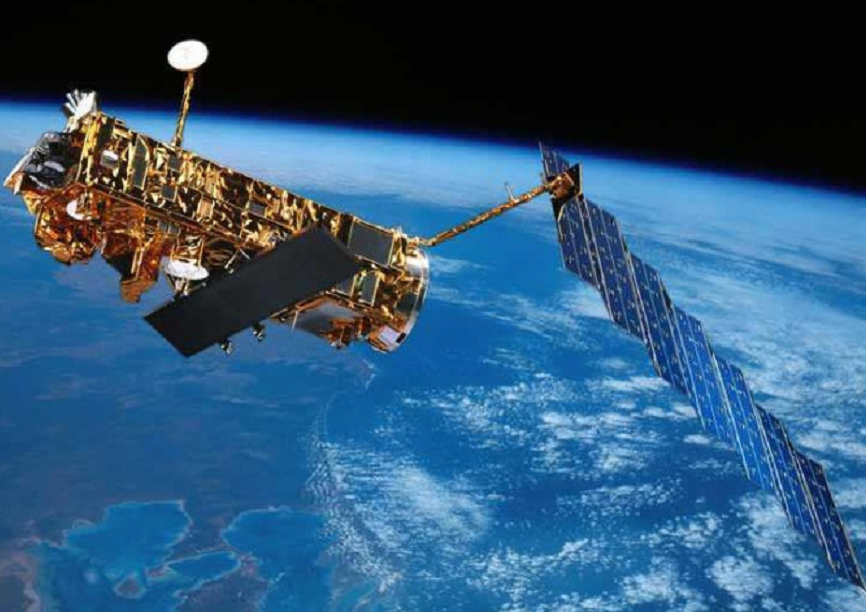}
%
%
\caption{Scientific ESA satellite ENVISAT (COSPAR no. 2002-009A). ESA lost contact with ENVISAT in April 2012. Is is the largest civilian non-functional satellite on low earth orbit and possible target of a future active debris removal mission. Image credit: ESA.}
\label{fig:envisat}       
\end{figure}

An example of a non-functional satellite can be seen in Figure \ref{fig:envisat}, shiwing the former ESA mission ENVISAT (international identification no. 2002-009A), which was successfully operating for about 10 years. In April 2012, however, ESA suddenly lost contact with ENVISAT and was not able to recover the connection. A few weeks later ENVISAT was declared as non-operational and the mission was terminated \cite{ESA201213}. Currently, ENVISAT is the heaviest civilian LEO satellite which is not operational and therefore it is a likely target for future active debris removal (ADR) missions \cite{LIOU11}.

%

Since the 1960s micrometer (dust) to centimeter (slag) sized particles were created during the burning process of $solid$ $rocket$ $motor$ (SRM) in space. Such residues are mostly composed of aluminium oxide, mixed with SRM liner material \cite{Horstman2009}. All space objects, such as satellites, rocket bodies and fragments are exposed to the rough space environment. Spaceweather effects, such as extreme ultraviolet (UV) radiation and interaction with atomic oxygen, cause erosion on the satellite surface which slowly ages and releases small particles into the space environment. Erosion has a strong influence mostly on painted surfaces and thermal protection materials. Because of its most common source this type of debris is denoted as $paint$ $flakes$. The large amount of small debris particles and meteoroids can create another population type of debris, called $ejecta$ \cite{SCHONBERG2001713}. After small particle impacts on painted surfaces, solar arrays or other type of spacecraft surface materials, impact craters are formed and some material is ejected into the space environment. Paint flakes and ejecta are usually sub-millimeter particles which do not pose a major risk to space missions, but they can cause degradation of surfaces and optical instruments. 

A specific type of debris are particles released from spacecraft with small additional velocities. Often they are caused by unknown mechanisms and referred to as $anomalous$ $debris$. One of the candidates for anomalous debris are multilayer insulations (MLI). MLI is material used as a thermal protection for sensitive systems of spacecraft (see Figure \ref{fig:envisat} for highly reflective gold-colored MLI). During breakup events, or under the influence of the space environment (impacts from small particles, extreme ultraviolet radiation) the MLI parts can be detached from the parent body. Typically, these objects have very high area-to-mass ratios A/M (HAMR) and highly reflective surfaces \cite{SCHILDKNECHT20081039}. Because of solar radiation (and the atmospheric drag for LEO passes), high A/M values have a strong influence on the dynamics of anomalous debris and may cause a drift away from the satellites' operational orbits \cite{Liou2005}. 

A unique population from a dynamic point of view are Temporarily Captured Orbiters (TCO) with an artificial origin. Those can be stages from former lunar missions which were gravitationally ejected into heliocentric orbits. Such objects have been observed several times in the last two decades once they were re-captured in the gravitational field of the earth and consequently have been mistaken for a Near Earth Asteroids (NEA) \cite{BUZZONI2019371}, \cite{ESA2013qw1}. These objects are on HEOs.

Parts of the space debris population do not have reproducing sources anymore, and therefore their populations are decreasing over time. Such population is cooling liquid droplets released during the Russian missions called Radar Ocean Reconnaissance Satellites (RORSAT). These droplets are consisting of a $sodium-potassium alloy$ (NaK) \cite{WIEDEMANN20111325}. Clusters of small, micrometer size needles created during the $Westford$ $Needles$ project in 1960s are another example a non-reproducing debris sources \cite{Klinkrad2006}.

\begin{figure}[b]
\sidecaption
\includegraphics[width=12.0cm,angle=0]{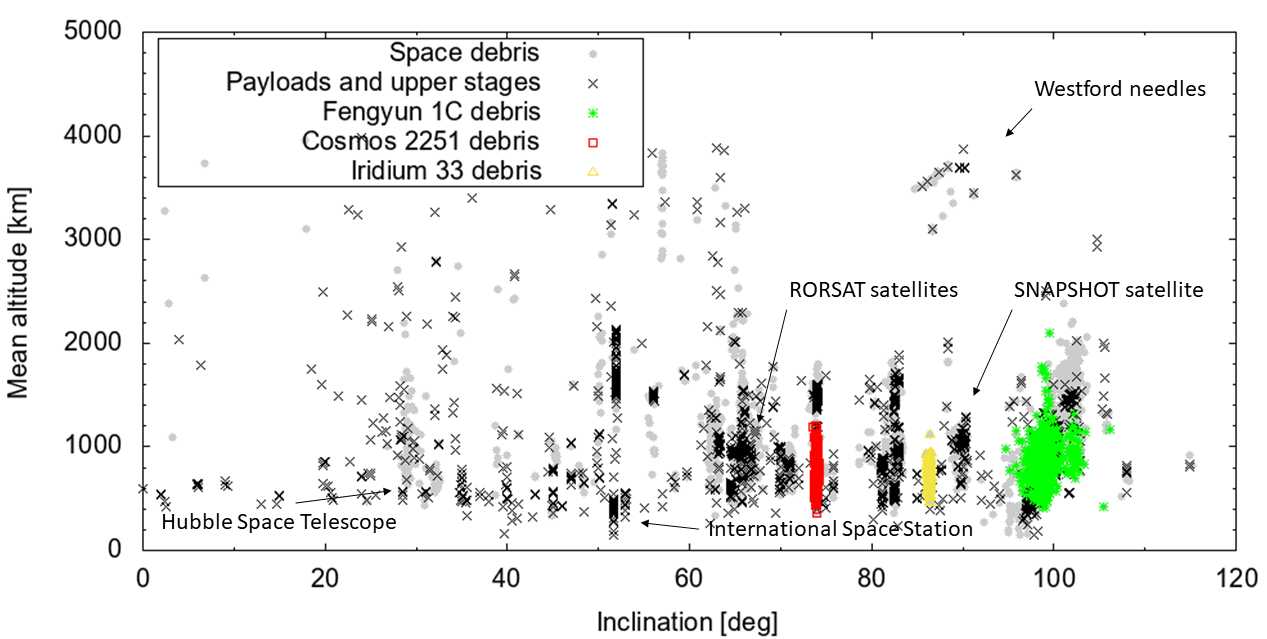}
%
%
\caption{Inclination vs mean altitude of catalogued object orbits with altitudes less than 4,000 km. It is possible to distinguish intact objects, such as payloads and rocket bodies, fragments from satellites Fengyun-1C (mean inclination of 99.0$^{\circ}$), Cosmos 2251 (mean inclination of 74.0$^{\circ}$), Iridium 33 (mean inclination of 86.4$^{\circ}$) and other catalogued debris. Marked are also orbits of the International Space Station (ISS), Hubble Space Telescope (HST), Russian RORSAT satellites, US SNAPSHOT satellites and Westford Needles clusters. Data were obtained from the public catalogue \cite{Spacetrack} (September 2019).}
\label{fig:type_vs_orbit}       
\end{figure}

A closer look at the LEO population from the perspective of object type versus orbit is plotted in Figure \ref{fig:type_vs_orbit}. Shown is the orbital inclination as a function of mean altitude above the earth's surface for different types of debris populations. The data were plotted by using the public catalogue.

%

\subsection{Population growth}
\label{subsec:spatial}
The future state of the space debris environment depends on several factors. The crucial one is the active debris removal (ADR) technology \cite{LIOU2010648}, \cite{WANG201822} which can help to remove potentially hazardous objects with high risk of fragmentation which eventually can cause a catastrophic cascade effect also known as Kessler syndrome \cite{KESSLER20041006}, \cite{Kessler2010}. Additionally, cheaper space technologies and new participants involved in  space industry such as commercial companies and academic institutions created a new form of satellite infrastructures like cubesats \cite{Matney2017} and broadband world-wide coverage mega-constellations \cite{DELPORTILLO2019123}, \cite{LEMAY2018445}.

\section{Measurement techniques}
\label{sec:techniques}

There are several ways how to measure thr space debris population, in terms of its size, mass, composition, and spatial distribution. Using optical telescopes help us to track objects of lower sizes from 10 to 50 cm on LEO and GEO orbits, respectively, and study their physical properties such as brightness and rotation. Thanks to radar measurements we are able to track and monitor LEO objects down to 10 cm in size. In-situ detectors and returned spacecraft hardware allow us to measure micrometer-sized debris and micrometeoroids, in terms of their flux, size and mass distributions. 

\subsection{Passive optical}
\label{subsec:optical}
The principle of passive optical space debris observations is based on collecting the sun light reflected from the object's surface. Therefore, to acquire such data several conditions need to be fulfilled: the object must be illuminated by the sun; it must be observed during the night; and the sky needs to be transparent, hence the weather conditions must be optimal.

Passive optical measurements provide so-called angle measurements and astrometric data, which are then used for orbit determination and object cataloguing. Additionally, surveys are performed to discover new objects and to gather new sampling data for statistical purposes. Light curve acquisitions help to monitor the rotational behaviour of objects, while multi-band photometry along with spectroscopy help to characterize the object and its surface properties. More about optical passive measurements can be found in Section \ref{sec:research}.

\subsection{Optical active, satellite laser ranging}
\label{subsec:slr}
Satellite Laser Ranging (SLR) systems, which are active optical systems, acquiring range and angular measurements. These observations are based on the photon reflection from the target. Contrary to passive optical measurements, a photon is emitted by the SLR system toward the object and reflected back toward the observer, where it is detected. The obtained information is the photon's duration of flight which can be recalculated to the range between the SLR system and the object. SLR measurements can be conducted during the day and night but they require good weather conditions. 

SLR systems are primarily used to accurately measure an object's range with accuracy up to few millimeters \cite{combrinck2010satellite}. Two types of SLR targets can be distinguished. [a] The object is equipped with retroreflector(s) (RR), a highly reflective device which reflects the light back to the source. These are cooperating targets which are used for scientific applications such as geodesy \cite{STRUGAREK2019417}. Cooperating targets are observable by any SLR station but their total number is limited (a few dozen worldwide) \cite{PEARLMAN2002135}; [b]The object is not equipped with retro-reflector. Such non-cooperative objects, are the vast majority, including all space debris. To observe these targets it is necessary to use much stronger lasers to get a statistically meaningful number of photons reflected back to the system. Those are usually experimental set-ups such as in \cite{KIRCHNER201321} or \cite{ZHANG2019691}. The final product of SLR measurements are range measurements which, apart from position information, can also provide information about the target's attitude state \cite{KUCHARSKI20142309}, \cite{PITTET20181121}.

\subsection{Radar}
\label{subsec:radar}
Radar measurements, in contrast with optical measurements, not limited by meteorological conditions and are executable during the whole day. Most radars are able to measure 2-way signal travel time, the Doppler shift between received and transmitted frequency, the azimuth angle $A$ and elevation angle $h$, and the received power and polarization changes in the radar pulse \cite{Klinkrad2006}. The parameter which describes the target's ability to reflect the radar energy is denoted as radar cross section (RCS). The $RCS$ depends on the target's size, the ratio of its characteristic length $Lc$ to the radar wavelength, and the target shape, orientation and material properties. The $RCS$ is expressed in decibel square meters (dBsm) and can be understood as equivalent to the visual magnitude used during the optical measurements \cite{Klinkrad2006}. The $RCS$ values for some debris objects can be retrieved from the public catalogue \cite{Spacetrack}.

There are several rsearch radar systems around the world participating in space surveillance and tracking. These are for example the US Goldstone and Haystack radars \cite{STOKELY2009364}, and European EISCAT \cite{MARKKANEN20051197} and TIRA radars \cite{MEHRHOLZ1997203}. 

\subsection{In-situ}
\label{subsec:insitu}
In-situ measurements mainly help to model the debris population, smaller than 1 mm. There are several models which use this information, for example ESA's MASTER model (Meteoroid and Space Debris Terrestrial Environment Reference) and NASA ORDEM model \cite{KRISKO2015204}. There were several in-situ experiments such as the Long Duration Exposure Facility (LDEF) \cite{MANDEVILLE199567} or Space Shuttles surfaces returned for impacts analyses \cite{HYDE2015246}. 

A good example of in-situ measurement is the LDEF mission. The mission lasted from April 1984 till January 1990, when the satellite was exposed to the space environment in LEO. The purpose of the LDEF mission was mapping the micrometeoroid and space debris (MMOD) environment, testing various materials (such as plastics and glass) and their suitability for space missions, and study effects of radiation and other space hazard. The satellite was located in LEO (mean altitude about 458 km and inclination around 28.5$^{\circ}$) and exposed to the space environment for more than 5.75 years \cite{MCBRIDE1995757}. There were 86 trays installed on board of LDEF with 57 different experiments and a total of exposed area of 130 m$^{2}$. The post mission examination of the thick and thin targets was performed to get the size, mass and flux distribution of the MMOD population \cite{MCDONNELL199825}.  

\section{Research with passive optical telescopes}
\label{sec:research}

This section discusses different research performed with passive optical telescopes. It covers the survey and cataloguing, light curve processing, color photometry and spectroscopy.

\subsection{Survey, astrometry and cataloguing}
\label{subsec:surveys}
Survey observations serve to discover new objects, either for statistical population modeling, or for tracking and cataloguing \cite{MOLOTOV20081022}, \cite{WEIGEL2017531}. Space agencies such as NASA and ESA are monitoring the MEO and GEO regions to obtain statistical information about faint non-catalogued objects \cite{SCHILDKNECHT2008119}, \cite{SILHA2017181}, \cite{ABERCROMBY2009103}.

Since 2001 NASA uses its Michigan Orbital Debris Telescope (MODEST) to scan the GEO region. MODEST is a 0.6-meter Curtis-Schmidt telescope equipped with the charged coupled device (CCD) camera with a 1.3 $\times$ 1.3 degrees field of view \cite{ABERCROMBY2009103}, \cite{Seitzer2011}. The telescope is located at the Cerro Tololo Inter-American Observatory (CTIO) in Chile. The acquired observations are processed by internal Image Reduction and Analysis Facility (IRAF) routines for photometry and astrometry, and the obtained results are compared with the public catalogue \cite{Spacetrack}. This helps to distinguish correlated targets (CT, i.e. objects have been correlated with the public catalogue) from un-correlated targets (UCT). An additional information from the data is the brightness (apparent magnitude)  $m_{app}$ of the object, its size $L_{c}$ by using some assumptions on the shape and albedo, and a preliminary subset of orbital elements, namely the orbital inclination $i$, right ascension of ascending node $\Omega$ and mean motion $n$. 


There were usually from 20 to 40 nights of observations per year, each split into three observations runs. The basic observation strategy was to choose equatorial coordinates for the field of view, with right ascension $\alpha$ and declination $\delta$ close to the Earth's shadow (or anti-solar point) to maximize the detection rate. Once the object is observed under small phase angles ($\varphi$ close to 0$^{\circ}$) it maximizes its reflection of the sunlight toward the observer. $\varphi$ is defined as the angle between the Sun-object and object-observer directions. The chosen field of view with the given $\alpha$ and $\delta$ was tracked during the whole night using a broad R filter centered at 630 nm, with a the full width at half maximum of 200 nm. The exposure time was set to 5 s, which led to a signal-to-noise ratio of 10 for objects of 18$^{th}$ magnitude. During the observations the time delay integration (TDI) method was used, where the charge on the CCD is shifted opposite to the sidereal rate. Hence, the stars on the exposure images are displayed as streaks and objects located on GEO are displayed as points or small streaks.

Once the apparent magnitude $m_{app}$ is obtained (see Section \ref{subsec:color_phot}) it is possible to get a characteristic length $L_c$ of the object, which characterizes the size of asymmetric debris fragments. It is defined as the average of three orthogonal dimensions X, Y and Z, where X is the longest dimension of the object, Y is the longest dimension perpendicular to the X axis, and Z is the longest dimension perpendicular to the other axes \cite{HANADA2009558}:

\begin{equation}
  \label{char_length}
  L_c = \frac{X+Y+Z}{3}.
\end{equation}

To get $L_c$ from $m_{app}$ we use following formula assuming the object is a Lambertian sphere \cite{Mulrooney2008}:

\begin{equation}
  \label{eq_sigma_per}
  L_c = 10^{(m_{sun}-m_{app})/5}{\frac{R}{\sqrt{A}}}\sqrt{\frac{6\pi}{sin{\phi}+(\pi-\phi)cos{\phi}}}, \quad
\end{equation}

or

\begin{equation}
  \label{eq_sigma_per2}
  L_c = 10^{(m_{sun}-m_{abs})/5}{\frac{R}{\sqrt{A}}}\sqrt{6}, \quad
\end{equation}

where the variable $Lc$ [km] is the diameter of a diffusely reflecting Lambertian sphere, $R$ is the observer-satellite distance [km], $m_{sun}$ is the apparent magnitude of the sun, $\varphi$ is the solar phase angle [rad], $A$ is the Bond albedo, and $m_{abs}$ is the absolute magnitude of object. $m_{abs}$ is defined as the magnitude of the object corrected for the phase angle ($\varphi$ = 0$^{\circ}$). The magnitude $m_{abs}$ can be determined as follows:

\begin{equation}
  \label{mag_abs_app}
  m_{abs} = m_{app} - 5 log(\frac{\pi}{sin{\phi}+(\pi-\phi)cos{\phi}}).
\end{equation}

\begin{figure}[b]
\sidecaption
\includegraphics[width=6.5cm,angle=0]{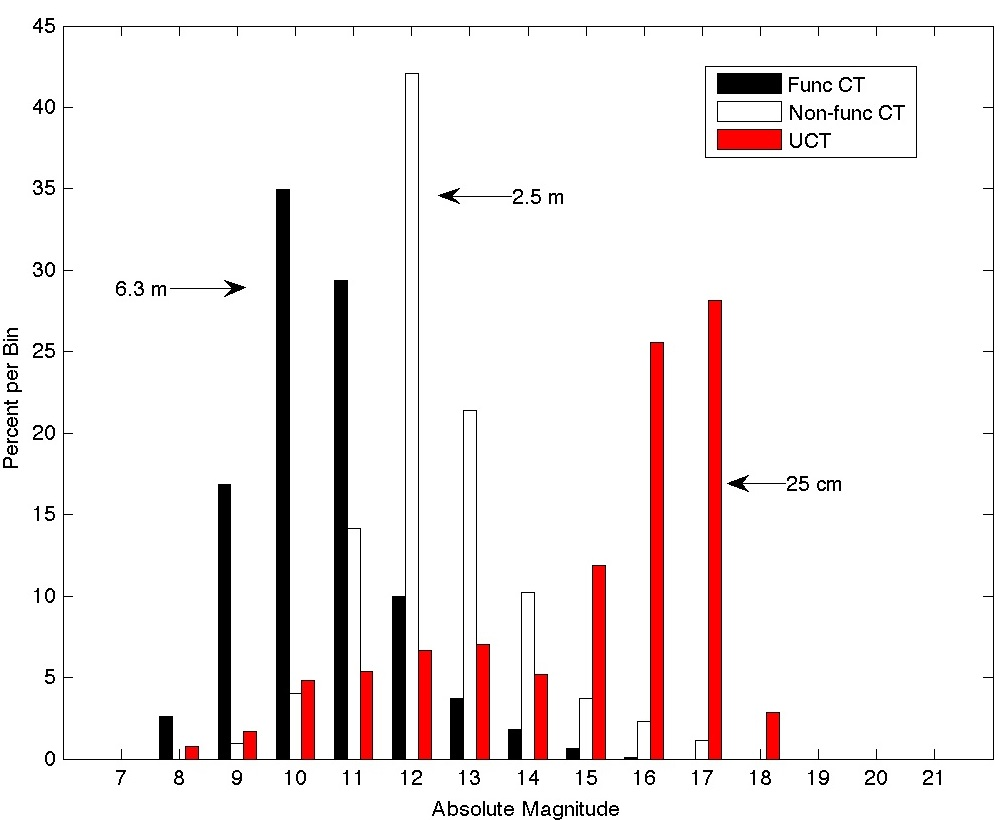}
%
%
\caption{The absolute magnitude distribution of 3,143 objects observed by the MODEST telescope during the years 2007, 2008 and 2009. Objects are separated into functional CTs, non-functional CTs and UCTs. Corresponding object sizes were calculated assuming a diffuse Lambertian reflectaion with albedo 0.175. Figure taken from \cite{Seitzer2011}.}
\label{fig:modest}       
\end{figure}

According to \cite{Mulrooney2008} the value A = 0.175 can be adopted as a transformation albedo to calculate the object size from equations \ref{eq_sigma_per} and \ref{eq_sigma_per2}. The absolute magnitude distribution of 3,143 objects which were observed during MODEST observation runs between 2007 to 2009 can be seen in Figure \ref{fig:modest} \cite{Seitzer2011}. Objects were on GEO and are separated into the functional CTs, UCTs and non-functional CTs, which are objects that are freely drifting in the north-south direction and east-west direction. The peak for the functional CTs is at $m_{abs}$ = 10.5 mag which under previous assumptions (see previous paragraph) can be interpreted as an object size of $L_c$ = 6.3 m. For non-functional objects the peak is at $m_{abs}$ = 12.5 mag, with corresponding $L_c$ = 2.5 m. The peak of the UCTs is an artificial one due to the limited detection capabilities of the telescope, and it can be assumed that the population continues rising with decreasing object size. The faintest object detected in 2007 - 2009 had a magnitude of $m_{abs}$ = 18.3 mag which corresponds to $L_c$ = 17 cm.

%

Once the object is discovered and sufficient follow-up observations are performed (at least two), its orbit can be determined and the object can be catalogued. There are several different perturbation forces affecting the dynamics of objects orbiting the earth. They need to be considered during the orbit determination and propagation process. These are gravitational effects such as earth's spherical harmonics and perturbations from the sun and moon, and solar radiation pressure and atmospheric drag. Thrust forces can also play a role. They are originating from the object but are not usually accounted for during routine orbit determination/prediction processes. There are several dynamical models available which deal with the satellite's orbital dynamics, e.g., described in \cite{BEUTLER2005}, \cite{Montenbruck2000}.

To get accurate orbits, highly accurate astrometric data need to be acquired. This specially goes for space debris objects which can reach apparent angular velocities up to few degrees per second. In this case even a small time error at the level of milliseconds can lead to an astrometric error of a few arc-seconds. For that reason optical sensors need to be monitored for possible time biases and inaccuracies. There are methods to perform such analysis, e.g., described in \cite{JILETE2019}.

\subsection{Photometry, light curves}
\label{subsec:ligh_curve}
Light curves carry an extensive set of information about the object's dynamical and physical properties. The brightness variation present in the data is directly related to the object's rotation and the mutual geometry between the sun, target and observer. The shape of the light curve is directly influenced by the sampling used during the data acquisition. Object's shape, its reflectivity properties expressed via the albedo, surface properties and the aspect angles, and the viewing angle from the perspective of the observer. There are many applications of light curves. Information related to the attitude determination \cite{WILLIAMS1979885}, \cite{SANTONI2013701} can directly be applied to the ADR problem \cite{LIOU2010648}, \cite{FORSHAW2017326}, \cite{WANG201822}. Regular monitoring of the rotation change over time helps to model forces influencing the object's dynamics, such as the electromagnetic field, atmospheric drag, solar radiation, and internal processes (fuel sloshing, outgassing) \cite{WERTZ1978}, \cite{EARL2014}. The shape estimation from light curves is quite often used in the minor planets domain \cite{KAASALAINEN200124}, \cite{CROWELL2017254}, and currently it is also being exploited for space debris objects \cite{Linares2012}, \cite{Bradley2014LIGHTCURVEIF}.

There are several different approaches and methods how to extract the frequency or apparent (synodic) period from the light curves \cite{GRAHAM2013}, \cite{PAUNZEN2016}, including visual inspection \cite{EARL2014} for simpler cases. The most common methods are Fourier-based methods. These are for example Fast Fourier Transforms (FFT), Discrete Fourier Transforms \cite{DEEMING1975}, Lomb-Scargle \cite{SCARGLE1982} or Welch periodogram analyses \cite{WELCH1967}. Frequently used are string-length methods \cite{PAUNZEN2016}. Their main approach is to fold the series into the test period, where the resulting folded light curve, phase diagram, is further analyzed. Those methods are more robust, and applicable to non-equally sampled data. Just to mention a few, there is epoch folding \cite{LARRSON1996}, wavelet analysis \cite{Astafyeva1996}, the Lafler-Kinman method \cite{LAFLER1965} and Phase Dispersion Minimization \cite{STELLINGWERF1978224}.  

\begin{figure}[b]
\sidecaption
\includegraphics[width=7.2cm,angle=0]{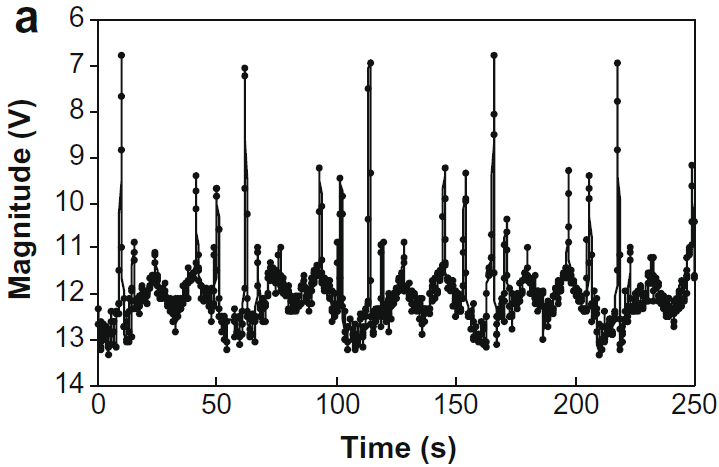}
%
%
\caption{The original light curve of the spacecraft Raduga 14 (1984-016A). Figure taken from \cite{PAPUSHEV20091416}.}
\label{fig:papushev_lc}       
\end{figure}

\begin{figure}[b]
\sidecaption
\includegraphics[width=7.2cm,angle=0]{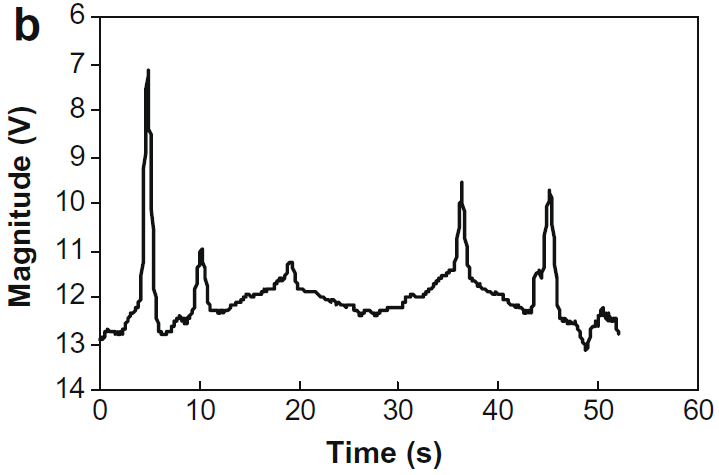}
%
%
\caption{The averaged light curve of the spacecraft Raduga 14 (1984-016A) developed by the Lafler-Kinman algorithm. Figure taken from \cite{PAPUSHEV20091416}.}
\label{fig:papushev_lk}       
\end{figure}

Between the years 1987 and 2004 the authors of \cite{PAPUSHEV20091416} acquired photometric data of 20 GEO satellites, where the majority were satellites of type Gorizont, Raduga, Ekran and Geizer. The authors used a two-mirror Cassegrain telescope with 0.5 m aperture equipped with a photoelectrical photometer setup, with a fast photometry mode, oparated at the Sayan observatory, Russia. This system was able to acquire data up to 1 kHz with a time accuracy of 1 ms. To extract the rotation periods the authors used FFT, the Lafler-Kinman method, and wavelet analysis. An example of an acquired light curve and reconstructed phase diagram constructed by using the Lafler-Kinman algorithm is plotted in Figure \ref{fig:papushev_lc} and Figure \ref{fig:papushev_lk}, which show data for the non-functional spacecraft Raduga 14 (1984-016A). The light curve, as well the phase diagram contains several sharp peaks which are typical for rotating, box-wing-type spacecraft. For these specific data the obtained apparent rotation period was 53 s. 

%

%

Authors of \cite{PAPUSHEV20091416} monitored several objects over long periods of time. They provided information about the change of rotation as a function of time. According to the variation types the authors distinguished three types of behaviour for rotating objects: long-term systematic increase/decrease of rotation; step-like variations; and sporadic, anomalous increase in period over several days. The latter variation type has been observed for Raduga 14, for which data are plotted in Figure \ref{fig:papushev_lc} and Figure \ref{fig:papushev_lk}. 

Research published in \cite{PAPUSHEV20091416}, as well as \cite{DEPONTIEU1997229}, \cite{BINZ2014}, \cite{SILHA2018844} showed that debris, including upper stages and spacecraft, cover wide range of rotation rates, up to few revolutions per second.

\subsection{Color photometry}
\label{subsec:color_phot}
Color photometry is based on multi-band photometric measurements. The difference between captured brightness in different bands is depended on the object's surface properties such as material color, roughness, albedo, etc. Color photometry can help to distinguish material types of an object's surface by comparing the results with laboratory experiments \cite{COWARDIN2010}, \cite{BEDARD2016} or at least it can help us to categorize an object according to its color \cite{CARDONA2016514}, \cite{ZHAO20162269}. Additionally, one can use photometric data to monitor the aging of the material due to space weather effects \cite{JORGENSEN20041021} (see \ref{subsec:origins}).

The photometric bands are defined trough their filters. There are several filter types in astronomy which are commonly used. Those are for example the $UBVR_{c}I_{c}$ filters for the Johnson-Cousins standard system \cite{CARDONA2016514}, and more modern $u'g'r'i'z'$ filters of the Sloan standard system used during the extensive campaigns of the Sloan Digital Sky Survey (SDSS) system \cite{Fukugita1996}, \cite{LU20172501}. Transformations between those two systems can be found in \cite{castro2018transformation}. While the Johnson-Cousins system allows broader interval for specific filters which are overlapping, the Sloan system has a narrow interval which allows clearly distinguish between different bands. 

Working with $UBVR_{c}I_{c}$ filters is very convenient mostly thanks to the extensive work of \cite{landolt1992ubvri} and \cite{landolt2009ubvri}. Here, the authors measured dozens of stars located around the equator by using $UBVR_{c}I_{c}$ filters. These so-called Landolt standard stars are used till today for the transformation from instrumental magnitudes to the Johnson-Cousins standard system. The conversion has a following form \cite{Benson1998}:

\begin{equation}
  \label{eq_mag_app_ins}
  m_{app} = m_{ins} - Z - \kappa X - color, \quad
\end{equation}

where $m_{app}$ is the apparent magnitude of the object in a given filter, $m_{ins}$ is the measured instrumental magnitude, $Z$ is the photometric zero point between the standard and instrumental systems, $\kappa$ is the atmospheric extinction coefficient, $X$ is the airmass (which is a function of the zenith distance $z$ [deg]), and $color$ is the color term transformation for a given filter, hence color index in instrumental system. For a specific filter, e.g. a $V$ filter, the equation has the following form:

\begin{equation}
  \label{eq_mag_app_ins}
  m_{app,V} = m_{ins,V} - Z - \kappa X - C(B-V), \quad
\end{equation}

where $C$ is the color coefficient and $(B-V)$ is the color index. By observing the Landolt standard fields during photometric nights, the parameters $color$/$C$, $Z$ and $\kappa$ can be determined for a given optical system. 

\begin{figure}[b]
\sidecaption
\includegraphics[width=8cm,angle=0]{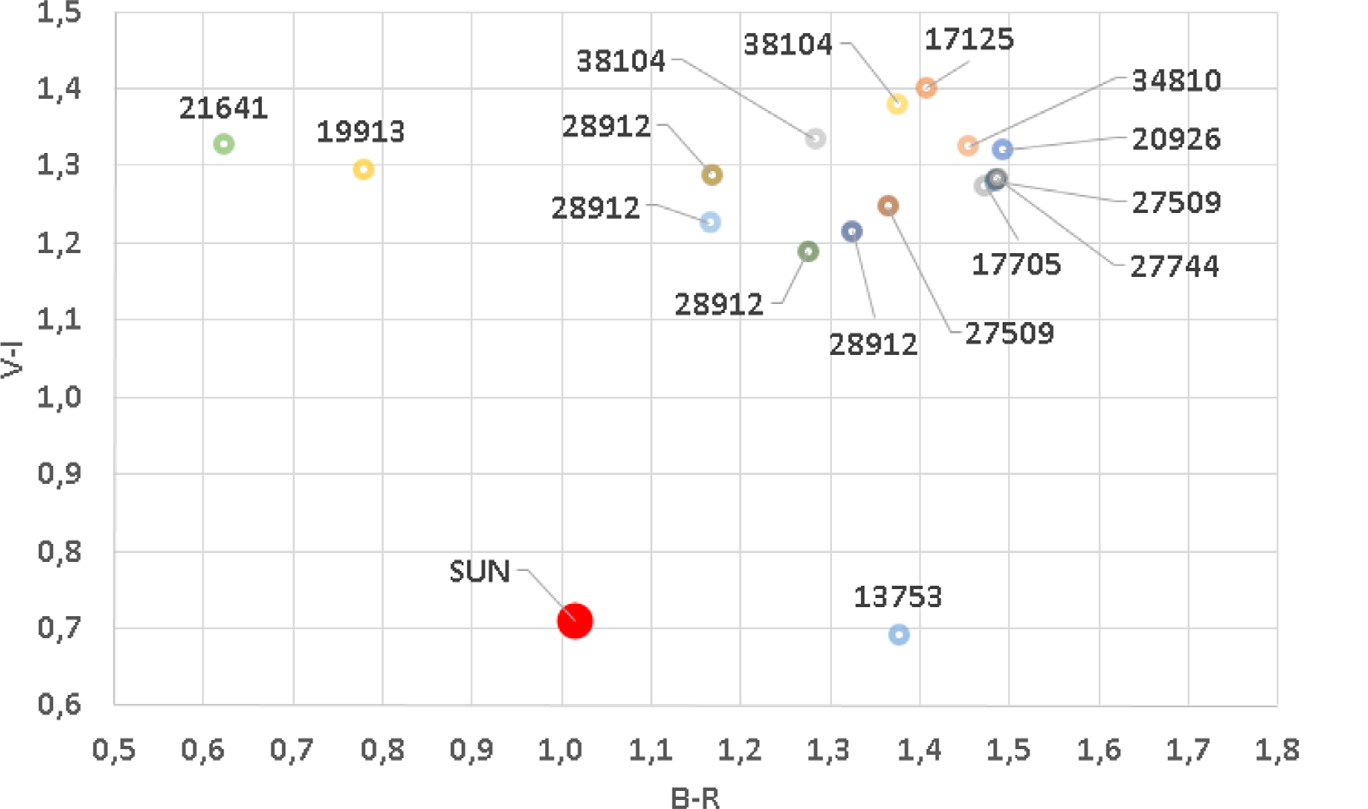}
%
%
\caption{Color indices $B - R$ vs $V - I$ obtained for thirteen GEO objects to investigate the surface properties and spaceweather effects. Figure taken from \cite{CARDONA2016514}.}
\label{fig:br_vs_vi}       
\end{figure}

In \cite{CARDONA2016514} the authors observed thirteen GEO objects in total, including seven upper stages, one non-operational spacecraft, two fragments and three operational spacecraft. The authors used Johnson-Cousins $BVR_{c}I_{c}$ filters to acquire photometric data. The Landolt stars were used for the transformation from the instrumental to the standard system. The acquired light curves and obtained values for color indices can be seen in Figure \ref{fig:br_vs_vi}, showing measured color index $B - R$ versus color index $V - I$, along with the color indices of the sun. The numbers in Figure \ref{fig:br_vs_vi} are the North American Aerospace Defense Command (NORAD) numbers which uniquely identify the object in the public catalogue. The results revealed that all objects (except one debris object) are redder than the sun in both indices. The authors also plotted color indices as a function of the object's launch date to investigate possible spaceweather effects. No visible trends have been observed.

The study \cite{CARDONA2016514} showed that the standard deviations of measurement points for one specific object are rather high. This is usually problematic for space debris objects where rotation rates are quite high (see Section \ref{subsec:ligh_curve}), and therefore the acquired photometric points can be covering large or relative large fractions of one rotation or even several rotations of the object. Therefore, the knowledge of the rotation properties of debris objects is essential once such type of measurements is acquired. 

%

\subsection{Reflectance spectroscopy}
\label{subsec:spectra}
The reflectance spectroscopy is measuring solar light reflected from the target. Its final output is a light spectrum, which is the reflectance as a function of wavelength. Reflectance spectra can provide, similar to color photometry (see Section \ref{subsec:color_phot}), information about the material composition. It can be used to characterize the object \cite{JORGENSEN20041021}, \cite{VANANTI20172488}, \cite{DeMeulenaere2018} or it can reveal whether the object is artificial or not \cite{BUZZONI2019371}. Reflectance spectroscopy has been adapted to the space debris domain decades ago, and its results are often used for comparisons with the spectra obtained in laboratory experiments for commonly used space materials such as solar panels, MLI, alluminium alloys, white paint, etc. \cite{engelhart2017space}, \cite{Bengtson2018}. 

\begin{figure}[b]
\sidecaption
\includegraphics[width=7.2cm,angle=0]{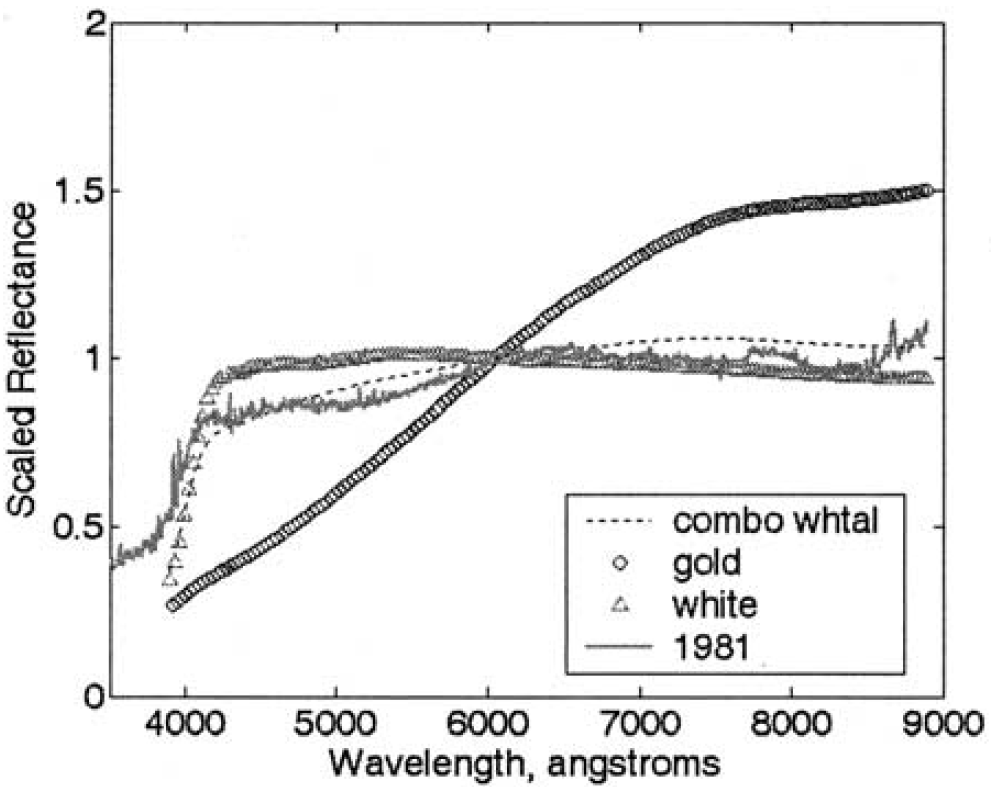}
%
%
\caption{Normalized reflectance spectra of an upper stage launched in the year 1981. Plotted are also spectra for materials such as "white paint", "degraded white paint, "gold", and their combination "combo whtal" (with an assumed ratio between "white" and "gold" paint to be 90~\% to 10~\%). Figure taken from \cite{JORGENSEN20041021}.}
\label{fig:spec_mat_type}       
\end{figure}

In \cite{JORGENSEN20041021} the authors acquired spectral data for various LEO and GEO rocket bodies and spacecraft within two campaigns: one concluded in 1999, and one performed in 2001. These data were acquired by the 1.6-m telescope located on at Maui, Hawaii, which is equipped with a spectrometer of three selectable gratings. The authors focused on three primary goals. The first goal was to compare the measured spectra with laboratory experiments to determine surface compositions. In Figure \ref{fig:spec_mat_type} reflectance spectrum is plotted for an upper stage launched in 1981. For comparison, laboratory spectra for "white paint" and "white paint after degradation" and "gold" are provided. Additionally, a combination of two materials, "combo whtal", is shown, with assumed a ratio between "white paint" and "gold" of 90~\% to 10~\%. This is the combination for which the authors got a very good match with the measured spectra.

\begin{figure}[b]
\sidecaption
\includegraphics[width=7.2cm,angle=0]{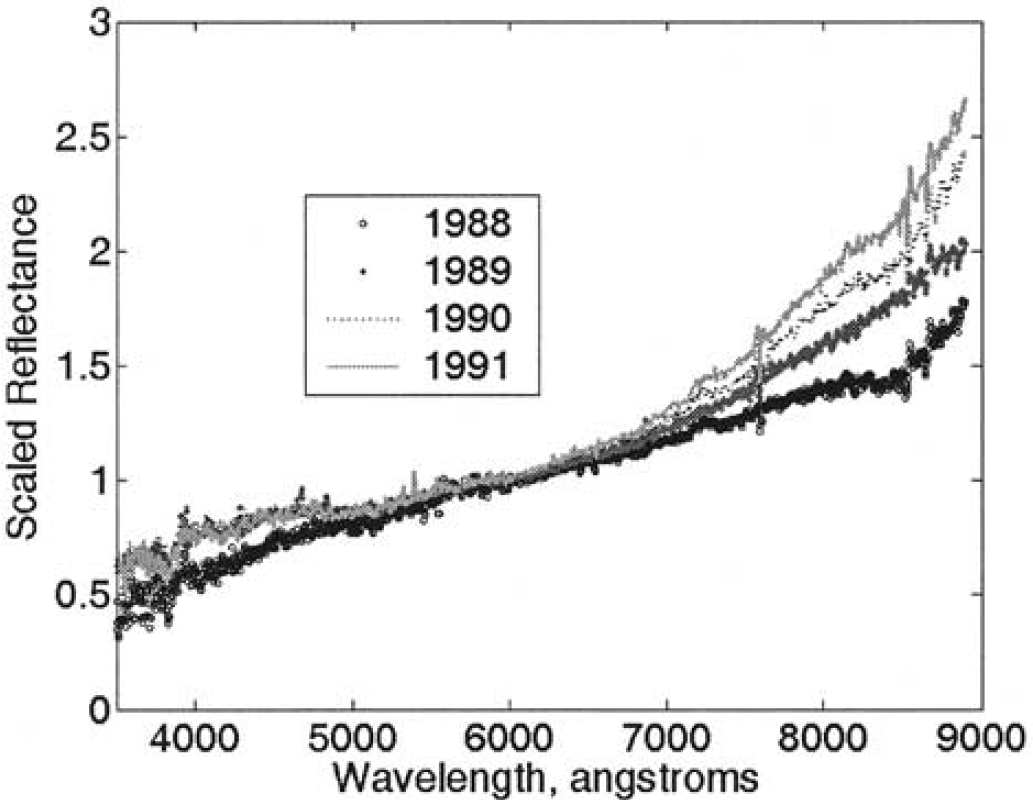}
%
%
\caption{Normalized reflectance spectra acquired for four different upper stages with similar properties, shape and surface material, but with different launching dates. This specific type of upper stage type spectra shows an inversly proportional effect for the increase with age. Figure taken from \cite{JORGENSEN20041021}.}
\label{fig:spec_aging}       
\end{figure}

As a secondly goal, the authors focused on an investigation of aging of the surface material. This could be achieved by observing the same type of upper stage with different launching dates. The results are plotted in Figure \ref{fig:spec_aging}, where reflectance spectra for four different upper stages of the same type (shape and surface material) are provided. The observed spectra showed an inversely proportional effect to the increase in age.  
As a third goal the authors also compared spectra between different types of objects, e.g., spacecraft versus upper stage. Spectra revealed that the most of the difference is in the blue part of the spectra, where satellites can reach much higher reflectance due to the presence of solar panels. 

%

%

\section{Conclusions}

Optical measurements play a strong role in space debris research. They are essential and irreplaceable for the tracking and cataloguing of objects on higher orbits such as geosynchronous and highly elliptical orbits. Typical astronomical methods such as sky surveys, photometry, or spectroscopy provide information about debris origins and creation mechanisms. The space debris population is constantly increasing and continuous improvements in optical data acquisition and processing can help to effectively deal with this problem. 

\biblstarthook{}

    \bibliographystyle{ieeetr}


\end{document}